# Capturing Financial markets to apply Deep Reinforcement Learning

Souradeep Chakraborty*

BITS Pilani University,

K.K. Birla Goa Campus

f20170170@goa.bits-pilani.ac.in

*while working under Dr. Subhamoy Maitra, at the Applied Statistics Unit, ISI Calcutta

**Abstract**

In this paper we explore the usage of deep reinforcement learning algorithms to automatically generate consistently profitable, robust, uncorrelated trading signals in any general financial market. In order to do this, we present a novel Markov decision process (MDP) model to capture the financial trading markets. We review and propose various modifications to existing approaches and explore different techniques like the usage of *technical* indicators, to succinctly capture the market dynamics to model the markets. We then go on to use deep reinforcement learning to enable the agent (the algorithm) to learn how to take profitable trades in any market on its own, while suggesting various methodology changes and leveraging the unique representation of the FMDP (financial MDP) to tackle the primary challenges faced in similar works. Through our experimentation results, we go on to show that our model could be easily extended to two very different financial markets and generates a positively robust performance in all conducted experiments.

# Introduction

## 1.1 Motivation

Since the early 1990s, efforts have been made to automatically generate trades using data and computation which consistently beat a benchmark and generate continuous positive returns with minimal risk. The goal started to shift from "learning how to win in financial markets" to "create an algorithm that can learn on its own, how to win in financial markets". As such, two major challenges, (apart from numerous others) have existed in this era of algorithmic/automated trading – existence of clean, reliable (or at least easily modifiable, using simple statistical transforms, to be reliable) data, and an effective computation algorithm that generates consistently (ideal) profitable or probabilistically more profitable trades.

With the passage of time and the understanding of the need for clean, easily available, free datasets led to the evolution of trusted, complete datasets. As such, one of the most secure datasets (in terms of reliability and completeness) is the Forex data, i.e. time-series price data (how much each currency-conversion, like the USD-INR, costed on a particular day). An interesting thing to note is that this data is dynamic, unlike say an image which represents static features, and hence one can appreciate the need for an *online* algorithm that can work dynamically and optimize itself with each additional test case.

The advent of *machine learning, deep learning* and *reinforcement learning* enabled us to come up with various algorithms that could solve complex tasks like image recognition completely automatically. This motivated researchers and financial institutions to try to come up with a machine/deep learning framework for trading. As such, several years of research was spent into suitably modeling the financial trading task, mostly failing. (Lopez de Prado 2018) discusses the primary reasons machine learning hedge funds have failed in the past.

The 2000s saw major advances in deep learning and many hybrid algorithms using core principles of deep learning and reinforcement learning were combined to finally come up with an effective algorithm that can be suitably modelled for trading – deep reinforcement learning. (Arulkumaran et al., 2017) is a brief survey of DRL.

The key area of application of most reinforcement learning algorithms are games – wherein the algorithm repeatedly tries to choose the best decision given a set of decisions. This decision-making process can be modelled using a Markov Decision Process. A brief survey of applications of MDPs is mentioned here (White 1993). This MDP is then solved using Q-learning, wherein the best action to take at each state of the environment is learnt. DRL is applied to this learning procedure to select an optimal action for each state-action pair in the MDP. DRL is especially beneficial for Q-learning in scenarios where value iterative processes like random exploration are infeasible due to the vastness of data or data that heavily depends on time – both of which are characteristics of financial datasets, thus making them an excellent candidate to be modeled using an MDP.

Thinking along similar lines, one can model the financial trading problem to be an MDP. While a lot of research exists on deep reinforcement learning as solutions to MDPs, deep reinforcement learning as an application to financial trading is a relatively new research area and as such, limited research exists on the topic. In this work we present a novel way to model the financial markets as an MDP and propose a completely online deep reinforcement learning algorithm to generate trades.

## 1.2  Introduction to the Financial MDP (FMDP)

This section assumes preliminary knowledge of Reinforcement Learning and Markov Decision Processes. For a brief review of these topics refer to (Kaelbling, Littman, and Moore 1996) and (White 1993).

The financial data environment, can be thought of to be highly dependent on time, to such an extent that it can be thought of as a function of time. As such, this property can be utilized extensively for working on an online-algorithm for coming up with trades.

Coincidentally, *Markovian processes* are defined to capture the entire past data and define the entire history of the problem in just the current state of the agent. When a reinforcement learning problem satisfies the Markovian property, the Transition function, (or Probability to go from one state to another) satisfies the above condition. Formally, this can be mathematically defined as below:

$$P(s_{t+1} = s', r_{t+1} = r | s_t, a_t, r_t \ldots, r_1, s_0, a_0) = P(s_{t+1} = s', r_{t+1} = r | s_t, a_t) \tag{1}$$

As such, we can appreciate how the financial environment fits into an MDP quite well. Consider for example, a company whose annual report just released, and that the company had suffered heavy losses in the last year. No matter the history or goodwill of the company, this news will negatively impact its stock price. This heavy dependence on newer events (and hence, time) helps us grasp how going from state to state in a financial environment would be most heavily dependent on the current state rather than past states.

## 1.3 Primary Challenges

In this section we present the major challenges faced while trying to apply deep reinforcement learning to financial markets.

Gathering data for such a sophisticated yet generalized task is the main challenge faced by researchers trying to model financial markets. High quality data is needed for such a task, and further this data needs to be quite exhaustive for the agent to learn using conventional deep reinforcement learning algorithms. Moreover, we also face the "*curse of dimensionality*" (Verleysen and François 2005) because of the exhaustive nature of the required data, which makes action exploration a challenging and computationally expensive task for the agent.

## 1.4 Expected Contributions

This work aims to explore the usage of simple mathematical modifications (dubbed *technical indicators*) to extrapolate easy-to-acquire data points (instead of high-quality OHCLV, or multi-instrument data). We expect these *indicators*, to efficiently capture the market dynamics as well as specific-instrument movements. This allows us to achieve a much easier state space market representation for the Markov decision model.

Further, the other aim is to finally design a deep reinforcement leaning agent that can learn optimal robust, profitable and uncorrelated strategies in any general financial market.

## 2 Methodology

This work aims to build upon the existing models applying deep reinforcement learning to financial markets, providing justified modifications. We show that our model is computationally less expensive to train (through experimental and theoretical analyses), and hence less exposed to latency. To achieve this improvement, we use a novel way to describe the Financial Markov Decision Process (FMDP) whose optimal policy is then found using deep reinforcement learning. This FMDP is modeled in such a way that it tackles the issue of data availability and action exploration. We then go on to experimentally demonstrate the improvements of our model (in section 3) and present results in section 3.3, utilizing financial mathematical ratios as metrics of evaluation

In this section we present our methodology, comparing existing literature while justifying proposed changes.

## 2.1 Describing the Financial MDP (FMDP)

Recall that every MDP with an agent has three basic components – the state space (the different states that the agent can be in), the action space (the different actions the agent can take in the different states) and the reward space (the reward that the agent gets corresponding to each action in each state). In order to model an MDP in financial trading, we choose the agent to be the trading algorithm – with the aim of constantly generating consistent profits in the environment – which is the financial market corresponding to the commodity being traded (for example, the stock market in case of stocks, forex market in case of currency pairs, commodities market in case of crude oil etc.).

This kind of a generalized Financial Markov decision process is unique; as all previous related work like (Huang 2018) and (Xiong et al. 2018), had models specific to the markets like the forex market and stock market respectively. This generalization is achieved by defining a simple state-space that is able to capture the task definition without including market-specific information.

### 2.1.1 State Space

Our FMDP is defined in such a way that an agent trades on any one particular security (say Crude Oil, or EUR/USD currency pair, etc.) and can buy/sell contracts (ranging from 1 to a fixed maximum number of contracts) of the corresponding security. The FMDP is designed to perform in diverse markets and as such, our state space explores the usage of *technical indicators,* which are mathematical and statistical modifications to the price data of the instrument. Using technical indicators may allow us to capture the instrument's correlations with the underlying market succinctly without having to use market-specific features. The other advantage of this kind of an approach is that these technical indicators are either freely available or can be obtained by simply using the time-series price data of the security.

The technical indicators we have used to capture the behavior of the security to be traded are – MACD (Moving average convergence divergence (Anghel 2015)), RSI (Relative strength index (Bhargavi, Gumparthi, and Anith 2017)), Williams %R (a momentum indicator invented by Larry Williams, Dahlquist, 2011), Weighted Bar direction (a parameter that tells us the direction and importance of the candlestick (William and Jafari 2011) formed by assigning weights) and previous day High – Low range. These indicators are chosen because of their simplicity and popularity.

In contrast, other works in this area typically use a market feature comprising of OHLCV data wherein either the data is quite exhaustive or closing price data is used directly as a feature along with closing prices of related securities (like in (Huang 2018), (Xiong et al. 2018), (Liang 2018) etc.).

Our representation of the state space not only makes it easy to port to other markets, but also captures intermarket relations like momentum, trend-reversal, etc as supported by (Lorenzoni et al. 2007) without having to rely on high-quality data (like OHCLV, which is not as easy to procure, beyond stock markets) or making the state-space too complicated.

The entire state space is thus majorly divided into two sub-parts:

- **Position State** – this is a 3D vector of the form $[L, S, PnL]$ where $L$ represents the number of contracts currently bought (number of long contracts in our holdings,) $S$ represents the number of contracts currently sold (number of short contracts in our holdings) and $PnL$ represents the corresponding profit or loss based on the current position.

    We provide further details regarding the Position State while discussing the Action Space in section 2.1.2.

- **Market features** – we utilize the 5 technical indicators, along with the timestamp (encoded to capture date-time), defined above to represent the market features. These features are crafted such that the models can extract meaningful information specific to the security and are scaled down to be ranging between 0.1 to 1 using a `MinMaxScaler`. (To normalize the data as explored by (Patro and Sahu 2015)) We also organize these indicators into time-series grouped according to timesteps.

    Intuitively, we can understand how this kind of a definition of market features make sense. Our agent is defined to mimic a real-world trader, and real-world traders seldom use vanilla price action data, and often work with technical indicators to study the underlying correlations more closely and accurately.

    Research supporting the validity and power of technical analysis in financial trading can be found in the works of (Lorenzoni et al. 2007) and (Hegde 2017).

    We further recognize that such a definition of the market features also makes the state space much less complicated (due to lower dimensionality), thereby reducing complexity and training time.

### 2.1.2 Action Space

The action space is implemented as a single value which could be 0, 1, or 2, representing Hold, Buy and Sell signals respectively, each of which is explained shortly. If at any state, the agent decides to perform a certain action, then correspondingly, only the position state of the State space will be affected. This is supported by the *zero-market impact hypothesis*, which essentially states that the agent's action can never be significant enough to affect the market features. This important fact is used to establish that no action can directly correlate with changes in the state-space's market features, thereby making the problem a little less complex.

If the action generated is that of a Hold signal, then the positions of the previous timestamp are carried over and no change is done to the position space. Further, if the action is corresponding to a Buy signal, then one long contract is added to the position space, provided the number of long contracts in the position space is less than the maximum number of contracts we can buy.

The sell signals in the action space work in the same manner. We have also defined the Position Space in such a way that at a time, we can either have only long contracts or only short contracts. This means that either $L$ (for action = 1), or $S$ (for action = 2), or both are zero (for action = 0) at a given timestamp. This kind of a definition of the position space also helps us later on while exploring policy for Q-Learning as we can calculate the *long-term reward* of taking a decision to Buy or Sell the security.

To better simulate real-world trading, we have also included transaction costs and these are incorporated into the PnL. The Position space also has details of the immediate PnL due to the action taken. This is obtained by calculating the price difference between the two timesteps and subtracting appropriate transaction costs applicable to the long/short contracts as described by the other two values ($L$ and $S$) of the position space (described in detail in the next section).

### 2.1.3 Reward and Policy

As our task is to effectively maximize our returns, it makes sense to somehow incorporate the PnL itself as a reward for the agent corresponding to each action. Usually, in MDPs, the reward of each action in each state is pre-defined, and the agent learns to choose that action over time which maximizes the reward. This can be done by either maximizing the immediate reward that the agent receives at each state due to each action (makes sense if the rewards can be calculated for every action), or by maximizing the eventual long-term reward that can be obtained by following the best set of actions after the current action is performed (Q-Learning).

In our FMDP, we utilize both the immediate and long-term reward. We already discussed how we continually store the immediate reward (or immediate PnL) due to a certain action in the position state. For implementing the long-term PnL, we calculate the PnL and keep adding it until a change in position is observed. We explain this implementation using the following example:

```
Initial State (timestamp t) = [0,0,0]
```
-> No long contracts, no short contracts, no PnL.

```
Action generated by model = 1
```
-> indicating Buy signal.

```
Price difference (price at t+T - price at t) = +55
```

```
Transaction cost = +5
```

```
Immediate Reward = +55 - (+5) = +50
```

```
Then, Final State (timestamp t+T) = [1,0,+50]
```

Again,

```
Action generated by model = 0
```
-> indicating Hold signal.

```
Price difference (price at t+2*T - price at t+T) = +30
```

```
Transaction cost = 0
```

```
Immediate Reward = +30 - 0 = +30
```

```
Then, Final State (timestamp t+2*T) = [1,0,+30]
```

Again,

```
Action generated by model = 2 -> indicating Sell signal.

Price difference (price at t+3*T - price at t+2*T) = -10

Transaction cost = +5

Immediate Reward = -(-10) - (+5) = +5

Then, Final State (timestamp t+3*T) = [0,1,+5]

And Long-Term PnL due to Long decision at timestamp t = 50 + 30 = 80.
```

This kind of a definition of the reward function allows the agent to easily find the *optimal policy*. The policy is essentially the probability distribution of each action at each state and hence defines the behavior of the agent. To effectively learn the optimal policy (which is the policy that maximizes rewards) the agent utilizes the $Q_\pi(s_t, a_t)$, the action-value function, which is defined according to the Bellman Equation as follows:

$$Q_\pi(s_t, a_t) = E_{s_{t+1}}[r(s_t, a_t, s_{t+1}) + \gamma E_{a_{t+1} \sim \pi(s_{t+1})} Q_\pi(s_{t+1}, a_{t+1})]] \tag{2}$$

With Q-learning, we basically learn the optimal policy (denoted by $Q_\pi^*(s_t, a_t)$) by learning the environment by choosing the greedy action – which instead of maximizing the expected value of the action-value function, chooses the action which maximizes the action-value function itself. Thus,

$$Q_\pi^*(s_t, a_t) = E_{s_{t+1}}[r(s_t, a_t, s_{t+1}) + \gamma * max_{a_{t+1}}(Q_\pi(s_{t+1}, a_{t+1}))]] \tag{3}$$

We see how the optimal policy according to this definition has an immediate component (the reward received $r(s_t, a_t, s_{t+1})$) and a long-term component (discounted by a factor $\gamma$).

Thus, we see that our definition of the FMDP beautifully suits the Q-learning methodology.

## 2.2 Optimal Policy - Deep Reinforcement Learning

Usually, a simple Deep Q-network, which performs function approximation using deep neural networks to encode states in the value-function is used to find the optimal policy. Reinforcement learning is done when we further use an experience replay buffer to store transitions and update the model network. However, this kind of an approach fails to do well in financial reinforcement learning problems (Verleysen and François 2005) because of the large action space size in these MDPs. To tackle this challenge, the authors of (Xiong et al. 2018) use an actor-critic network which utilizes two deep neural networks – one to map actions to states and one to get the value of that action. However, such an approach is computationally expensive and we see that authors of (Huang 2018) use action-augmentation to define a special kind of loss function, stated below, for the policy which could be solved using a simple network because of the simplicity of the action-space and the way rewards were defined.

$$L(\theta) = E_{(s,a,r,s') \sim D}\left[\left\|r + \gamma Q_{\theta^-}(s', \arg max_{a'} Q_\theta(s', a')) - Q_\theta(s, a)\right\|^2\right] \tag{4}$$

$$\theta \leftarrow \theta - \alpha \nabla_\theta L(\theta), \tag{5}$$

where $Q_{\theta^-}$ denotes the target network.

We use the same action-augmented loss function as reproduced above, while recognizing that there still exists a need to incorporate simpler policy exploration techniques – which include both immediate and long-term rewards. Consequentially, our novel definition of the FMDP allows us to model a network that finds optimal policy based on both the immediate reward and long-term reward as defined in section 2.2. Both these rewards are remembered by the agent (till it exhausts its memory, at which point it is reinitialized) and this cumulative reward (addition of these two log-returns, since RL applications work better with log definitions of rewards owing to its additive nature) is then used in the experience replay module of the agent, wherein it fits a deep reinforcement learning based network to generate the state-action value function for choosing the optimal action, based on the current state and memory of the previous states and rewards.

## 2.3 Network Architecture

We use a deep reinforcement learning network which takes the state-space as an input to generate the optimal policy to be used by the agent. We use this to select the best action to be taken by the agent, given the state space.

Recall that our definition of the state-space had two key parts (as discussed in Section 2.2.1) – the position-state and the market-features. We thus utilize this modularity in our model, by feeding the position-state directly to a fully-connected layer, while we use a combination of two LSTMs to effectively flatten the market-features. These are then merged together and then fed into a combination of two fully connected layers to finally generate the prediction of the value-function for that state. We use a ReLU activation (Agarap 2018) for all the layers except the last fully connected layer, wherein we use a softmax activation (Bridle, 1990). The Adam optimizer is used for the model (Kingma and Ba 2014). A complete block diagram description of the model is shown below in Figure 1:

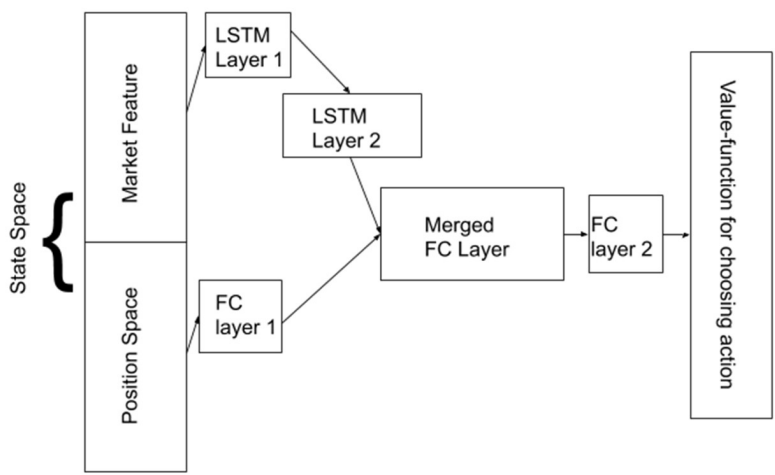

**Figure 1: Block diagram description of network to predict best action using to take at each state by generating the value-function given the state-space.**

As is evident, the network has been kept relatively simple to facilitate faster training.

## 2.4 Training

We utilize the training modifications originally proposed by (Huang 2018), to train the model to learn the optimal value function. Namely, we utilize a smaller replay memory (owing to the high dependency on more recent events in Financial markets) and we train the model every T timesteps (as this is beneficial to real-life trading since training can be done after market hours and the program can run with low latency). Further, instead of sampling a random sequence from the agent memory, we utilize the entire memory of the agent in the experience replay buffer. This helps us better capture long-term dependencies, that may have been missed because of the small replay memory. Thus, we maintain a kind of compromise between replay memory and sampling sequence.

### 2.4.1 Learning algorithm

We use the algorithm described below for learning the action-value function. We maintain a completely online algorithm, as this is more practical in trading applications, wherein it does not make sense to have the entire data from the beginning and performing a train-test-validation split, instead data comes in a sequential manner.

```
1.  Initialize T (positive integer, timesteps), recurrent    Q-network (θ), targe
t network (θ¯), dataset D,           environment E, learning rate τ
2.  Simulate environment E from dataset D
3.  Observe initial state s from env E
4.  for t = 1, T do:
5.      if rand(0,1) < epsilon:
6.          if epsilon > epsilon_min:
7.              epsilon*= epsilon_decay
8.          end if
9.          action = random action
10.     else:
11.         select action according to current model
12.     end if
13.     execute action and receive reward and next state
14.     store transition in memory buffer
15.     if memory of agent gets full:
16.         run experience replay
17.         store contents of memory in buffer
18.         clean memory
19.         fit Q-net model using equations (4) and (5)
20.     end if
21.     Update θ¯ <- (1 - τ) θ¯ + τ θ
22. end for
```

Using the initial rewards due to random actions, the model learns to take better actions to maximize these rewards ($\varepsilon$-greedy). This brings us to another important point – in the equation (4) used to fit the model, we use our special definition of reward (incorporating both immediate and long-term reward) instead of using the vanilla definition of reward as used in other reinforcement learning tasks.

## 2.5 Analysis of proposed architecture

In this section, we cover key methodology innovations and provide theoretical justifications as best as possible, before moving on to Section 3 to look at the Experimentation results, to prove the efficacy of our architecture.

# 3 Experiment and Performance Evaluation

## 3.1 Data Preprocessing

As our FMDP-DRL model is generalized, we can apply various datasets on our model. In our experiments, we have applied our model to Crude Oil and 9 different currency pairs in the Forex market (10 total instruments tested). We discuss in detail the results obtained from the Crude Oil dataset while listing the result details in tabular form for the currency pairs.

To prepare each dataset, the only raw data we need is the timestamp along with the price at that timestamp. This price data is easily available from online sources (like `TrueFX.com` for currency data). Once we have the price and timestamps, we can create the technical indicators using their mathematical expressions. For example, we have used MACD as one of the technical indicators. The mathematical expression for MACD is as follows:

$$MACD = 26 \ day \ EMA - 12 \ day \ EMA \tag{6}$$

where EMA represents the Exponential Moving Average of the price. Details regarding the technical indicators used can be found in section 2.1.1. We also used the `MinMaxScaler` module of `sklearn` to scale down all the market features discussed previously to a range of 0.1 to 1 to ensure data normalization.

As we are using a completely online algorithm for learning, we do not need to worry about a train/test/validation split.

## 3.2 Experimental Settings

In our experiment, we set a maximum holding size of 5 contracts. This means that if consecutive buy signals are generated for more than 5 days, they will be treated as hold signals. The same goes with sell signals. Further, we use a constant commission based on the number of contracts we are buying/selling. This commission is a hyperparameter and has an impact on the performance of the model. Intuitively, we can understand this by connecting that the lower the commission, the more the agent tends to prefer active trading, as opposed to preferring lesser number of trades in case of a high commission. As such, we chose a commission of 2$ per contract (a reasonable value considering brokerage charges worldwide). In our experiments we also found that as long as the commission value stayed within a certain range (below 5$ per contract), the number of trades were within a realistic range.

For the agent, we chose a Discount Factor ($\gamma$) of 0.8 and a learning rate of 0.001. Further, as defined in 2.4.1, we choose a value of 1.0 for epsilon, a decay of 0.995, and a minimum value of 0.01.

## 3.3 Back-testing and Results

We chose a period of November 2018 to February 2019 for back-testing our model. Before getting into the Results of the experiment, we show how our agent's actions are interpreted during back-testing. In Figure 2, we show the typical behavior of our agent.

```
Buy action generated. Current state: [ 1.   0.  -0.09]
Sell action generated. Current state: [2.   0.   0.41]
State after executing critical Sell action: [0.   1.   0.35] Long-term PnL: 0.410
Buy action generated. Current state: [0.   1.   0.35]
State after executing critical Buy action: [1.   0.   0.34] Long-term PnL: 0.3500
Buy action generated. Current state: [1.   0.   0.34]
```

**Figure 2: Typical behavior of our agent** (PnL represented in x$10^3$ form)

As we can see, a buy action was generated, when there was already a Long contract in our position space. We see that upon buying the additional Long contract, our position space now reflects the new PnL (0.41 x10^3) and 2 long contracts instead of one. At this point, we see a Sell action generated. We term this a *critical sell* as this action results in the calculation of the long-term PnL due to the previous Buy actions.

We analyze the result obtained by our model on the Crude Oil prices during this time-period in Figure 3.

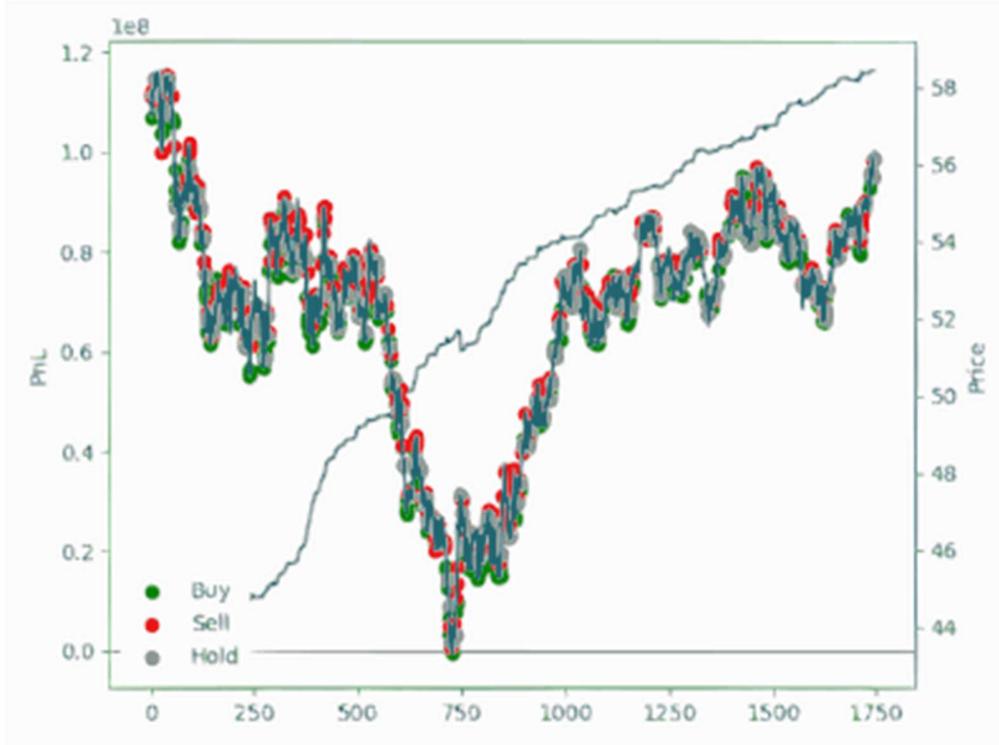

**Figure 3: Detailed plot of crude oil prices vs PnL of agent**

We see in Figure 3, that the agent performs robustly and provides substantial returns. Before presenting numerical results, we analyze Figure 3 with the focus on the actions taken. We see that the agent is able to utilize the information efficiently and take rational trades – learning how to buy/sell and hold at correct times and we also see how most of the actions taken aren't Buy or Sell actions, but rather Hold actions, which makes sense in real-life scenarios and is proven to work better because of lesser transaction costs and compounding effects of returns (*Buy-and-Hold* as discussed in (Shiryaev, Xu, and Zhou 2008)).

As our agent has no upper cap / fixed amount that it trades (rather a fixed maximum number of contracts), an absolute value of return is not a good metric for evaluation of our model performance because we trade on various different instruments with varying prices. As such, a better metric for evaluation is the annualized Sharpe ratio (Lo 2002) which combines returns and risk of investment together. Apart from the Sharpe ratio, we use two other metrics for evaluation of performance. These are the Win Ratio and Maximum Drawdown (Magdon-Ismail and Atiya 2004), both expressed as a %. Win Ratio is the ratio of the number of trades that generated positive PnL to the total number of trades, whereas, Maximum Drawdown (MDD) represents the downside risk and essentially represents the largest loss taken by the agent.

A summary of the performance of the model in various datasets can be found in Table 1.

| Instrument Name | Sharpe | Win Ratio | MDD |
| --- | --- | --- | --- |
| Crude Oil | 4.09 | 67.88% | -7.33% |
| EURUSD | 2.88 | 60.41% | -9.81% |
| GBPUSD | 2.57 | 61.29% | -10.63% |
| AUDUSD | 1.98 | 62.03% | -8.16% |
| GBPJPY | 3.55 | 59.62% | -9.47% |
| EURJPY | 2.23 | 61.31% | -5.33% |
| AUDJPY | 3.72 | 65.66% | -11.66% |
| CADJPY | 1.90 | 60.77% | -9.00% |
| NZDUSD | 3.57 | 63.82% | -6.39% |
| USDCAD | 2.01 | 59.23% | -8.83% |

Table 1: Summary of results of this work

We also reproduce the top performing results published in the works of (Huang 2018) and (Xiong et al. 2018) below in Table 2 and 3 respectively.

| Instrument Name | Sharpe | Win Ratio | MDD |
| --- | --- | --- | --- |
| AUDNZD | 5.7 | 63.20% | -1.21% |
| CHFJPY | 3.1 | 61.50% | -7.71% |
| GBPJPY | 2.9 | 60.80% | -7.73% |

Table 2: Top results of (Huang 2018)

Note that the format of the following table is different (as per (Xiong et al. 2018)).

| Algorithm / Strategy | Sharpe | Initial Capital | Annualized Return |
|---|---|---|---|
| DDPG (original work of authors) | 1.79 | 10,000 | 25.87% |
| Min-Variance (baseline used by authors) | 1.45 | 10,000 | 15.93% |
| DJIA (baseline used by authors) | 1.27 | 10,000 | 11.70% |

Table 3: Results of (Xiong et al. 2018)

We observe that our agent is quite robust in all the cases owing to a high Sharpe ratio. Moreover, we see that the Win Ratio of our agent is high which means our agent is accurately able to identify the directional movement of the instrument too – translating to a more robust trading strategy. Further, we observe the MDD to be within a comfortable range and hence we can say that there is not much downside risk associated with the strategies learnt by the agent.

# 4 Conclusion

In this work we establish a deep reinforcement learning algorithm to effectively learn the financial environment by modeling the task as a unique Markov decision process. We propose several changes to existing works in this area and support them by theoretical justifications by linking them with intuitive real-life financial scenarios and substantiate them using experimental results.

The primary methodology difference lies in the way the FMDP has been modeled. By defining a market-feature space that is independent of market-specific data, but still able to capture the overall market dynamics (Lorenzoni et al. 2007) using simple mathematical operations on the price data of a particular instrument, we expect to effectively reduce the state-space complexity while still maintaining robustness of information. (discussed in section 2.2.1)

Further, we introduce a simple action space, and use our novel definition of reward (incorporating both immediate and long-term reward based on a decision) to better suit the Q-learning methodology for solving the optimal policy (discussed in section 2.2.3). We then propose a deep reinforcement learning model to effectively find the optimal policy such that the agent is able to take trades that maximize performance.

Finally, due to the way we have discretely defined the two parts of the state-space - a market-feature which is essentially a time-series data-frame and a position-space, which is more like a discrete collection of data-points. We utilize this modularity in the state-space by using a recurrent neural network (two LSTM layers) on the time-series part of the state-space, (since we know RNNs work well with time-series data (as supported by the works of (Zhang et al. 2019) and (Balkin 1997)) while using dense networks on the position-space. If this modularity in the state-space were not maintained, we would not be able to apply different networks to learn the state-space features and thus would not capture the environment as well.

The next section summarizes the key achievements of this work. Further, in section 4.2, we look at future prospects for research in this area.

### 4.1 Achievements

- We propose a novel way to model the financial trading task as an MDP (FMDP). We make use of popular technical indicators to capture instrument specific market correlations. This makes our FMDP extendable to any financial trading task and reduces complexity in the state-space without a compromise in performance.

- We introduce a technique to model the reward function such that the agent is able to interpret both immediate and long-term effects of its actions and propose modifications to existing deep reinforcement learning algorithms to better suit the FMDP in particular, and the financial trading setting in general.

- We provide experimental results, (incorporating real-life trading constraints like commission per trade) for instruments traded in two different markets – commodities (crude oil) and forex (currency pairs) and achieve results that show that the strategies learnt by the agent are positive, uncorrelated and robust enough to be deployed in real-time.

### 4.2 Future Work

- Expanding to other trading scenarios like usage of a portfolio instead comprising of multiple instruments (instead of a single instrument), placing limit orders based on capital available for investment (rather than number of contracts), high-frequency trading etc.

- Apply and explore more sophisticated reinforcement learning techniques like those used in (Wang et al. 2018) and (Bellemare, Dabney, and Munos 2017), and the recent state-of-the-art Rainbow RL, which incorporates many different improvements to vanilla RL agents.

- Exploring other MDP models to succinctly capture markets. While we have proposed using technical indicators in place of high-quality data, there may be other ways to easily capture the financial market features (like fundamental data, in the case of stocks).

## 5 Acknowledgements and Declaration of Interest

### 5.1 Declaration of Interest

The author reports no conflicts of interest. The author alone is responsible for the content and writing of the paper.

## 5.2 Acknowledgements

This work would not have been possible without the constant support and guidance of Dr. Subhamoy Maitra, from the Applied Statistics Unit, ISI Calcutta. I am also grateful to several seniors and professors at BITS Pilani, Goa Campus, for exposing me to the world of research, computational and quantitative finance. Finally, I would like to thank my family, especially my parents, whose love and guidance shaped who I am today.